\documentclass[published]{JHEP}
\JHEP{01(1999)006}

\def\beq{\begin{equation}}
\def\eeq{\end{equation}}
\def\bea{\begin{eqnarray}}
\def\eea{\end{eqnarray}}
\def\bq{\begin{quote}}
\def\eq{\end{quote}}

\title{The large $N$ limit of QCD and the collective field of the Hitchin
 fibration}

\author{Marco Bochicchio \\
	INFN Sezione di Roma, Dipartimento di Fisica \\
	Universit\`a di Roma ``La Sapienza'' \\
	Piazzale Aldo Moro 2 , 00185 Roma  \\ 
	E-mail: \email{Marco.Bochicchio@roma1.infn.it}}

\abstract{By means of a certain exact non-abelian duality transformation, we show that
there is a natural embedding, dense in the sense of the distributions in the
large $N$ limit, of parabolic Higgs bundles of rank $N$ on a fiber
two-dimensional torus into the QCD functional integral, fiberwise over the 
base
two-dimensional torus of the trivial elliptic fibration on which the
four-dimensional theory is defined. 
The moduli space of parabolic Higgs bundles of rank $N$ is an integrable
Hamiltonian system, that admits a foliation by the moduli of holomorphic
line bundles over $N$-sheeted spectral covers (or, what is the same, over a 
space of $N$ gauge-invariant polynomials), the Hitchin fibration.
According to Hitchin, the Higgs bundles can be recovered from the
spectral covers and the line bundles.
If the $N$ invariant polynomials together with the abelian connection on 
the line bundles are chosen as the $N+1$ collective fields of the Hitchin
fibration, all the entropy of the functional integration over the moduli of 
the Higgs bundles is absorbed, in the large $N$ limit, into the Jacobian 
determinant of the change of variables to the collective fields of the Hitchin
fibration.
Hence, the large $N$ limit is dominated by the saddle-point of the effective 
action as in vector-like models.}

\received{October 9, 1998}
\accepted{January 8, 1999}

\begin{document}          

\section{Introduction}

The aim of this paper, that is a brief summary of a more extended one, is to 
gain control over the large $N$ limit of QCD in four dimensions~\cite{H}.

We consider QCD on a four-dimensional torus, that we think as a two-
dimensional fiber torus $T^{2}_f$, with complex coordinates $(z, \bar z)$,
trivially fibered over a two-dimensional base torus $T^{2}_b$, with complex
coordinates $(u, \bar u)$. 

The basic idea consists in changing variables, as in a fiberwise
duality transformation~\cite{B}, from the four-dimensional gauge
connection $(A_z,A_{\bar z},A_u,A_{\bar u})$ to the cotangent space of
unitary connections on the fiber torus, $T^*\cal{A}$, fiberwise over
the base torus, with coordinates $(A_z,A_{\bar z},\Psi_z,\Psi_{\bar
z})$.  $(\Psi_z,\Psi_{\bar z})$ are the components of a one-form,
$\Psi$, the Higgs current, on the fiber torus fiberwise, that label
the cotangent directions to the unitary connections $(A_z,A_{\bar
z})$.

$T^*\cal{A}$ is an infinite-dimensional hyper-Kahler manifold on which the 
gauge group, $SU(N)$, acts by Hamiltonian vector fields with respect to all
the three symplectic forms~\cite{H1,H2}. 

We then quotient, by the action of the gauge group, over a densely
chosen subset of $T^*\cal{A}$~\cite{B1}, in order to make the quotient a 
separable manifold.

The quotient turns out to be the inductive limit, with respect to the 
partial ordering defined by inclusion of any divisor, of the moduli space of
parabolic (with respect to a given divisor) $K(D)$ pairs~\cite{Y} on the fiber
torus fiberwise. This moduli space is a Kahler but not a 
hyper-Kahler manifold.
However, requiring the residue of the Higgs field to be nilpotent, a
hyper-Kahler reduction is obtained, corresponding to the moduli of 
parabolic Higgs bundles~\cite{Y,K}. The last one is a closed subspace
of the moduli of parabolic $K(D)$ pairs. The dimensions of the two mentioned 
moduli spaces are of order $N^2$. Therefore 
the two cases present equivalent difficulties from the point of view of solving 
the leading large $N$ limit. In fact, they differ only at sub-leading order 
$\frac{1}{N}$. We will not attempt to give a physical interpretation of this 
fact in this paper.

\pagebreak[3]

So far we have transformed, on a densely defined subset, the unitary
monodromies along four-dimensional curves, a local system with order
of $N^2$ degrees of freedom per point, into another local system, with 
the same order of $N^2$ degrees of freedom per point, the moduli of parabolic
$K(D)$ pairs or the moduli of
Higgs bundles fiberwise. Indeed, the last one is labelled
by the monodromies, with values in the complexified gauge group, around
arbitrary points on the fiber torus fiberwise~\cite{H1,S}.

We now make the last crucial step, that allows us to put the QCD partition 
function in the form of a vector-like model.

According to Hitchin~\cite{H3,N,M1,M2}, the moduli of $K(D)$ pairs or of Higgs 
bundles is an integrable 
Hamiltonian system, foliated by the moduli of line bundles over spectral 
branched coverings. These spectral covers are obtained from a basis of $N$ 
gauge-invariant polynomials in $\Psi_z$ via the characteristic equation.
Remarkably, all the information of the 
parabolic Higgs or $K(D)$
bundles on the fiber torus can be reconstructed from a spectral covering of the
torus and a line bundle on the cover~\cite{H3,N,M1,M2}.

We are thus recovering a rank $N$ local system on the fiber torus (with a 
parabolic divisor) from a rank $1$ local system (with a parabolic divisor) on
a $N$-sheeted bran\-ched covering of the fiber torus.

Yet, the last system is completely determined by a collection of
$N+1$ local fields, an abelian $U(1)$ gauge connection, that determines the
line bundle on the cover, and $N$ meromorphic differentials, holomorphic away 
from the 
divisor, that determine the spectral cover.

We call these fields the collective~\cite{J} fields of the Hitchin fibration.

Thus, by changing variables to the collective fields of the Hitchin
fibration, the QCD partition function looks like the one of a vector-like 
model.

As a consequence, the large $N$ limit is dominated by the saddle point 
of the effective action, that now includes the Jacobian determinant of the
change of variables from the moduli of parabolic Higgs bundles to the
collective field of the Hitchin fibration.

\section{The main formulae}

Our starting point is the $SU(N)$-$YM$ functional integral:
\bea
Z&=&\int \exp\left[- \frac{1}{4g^2} \int \sum_{\mu \nu} Tr\left( F_{\mu \nu}^2\right) d^4x\right] DA \,,
\nonumber \\ 
F_{\mu \nu}&=&
\partial_{\mu}A_{\nu}-\partial_{\nu}A_{\mu}+i [A_{\mu},A_{\nu}] \, . 
\eea  
The four-dimensional space-time is chosen to be a four-dimensional 
Euclidean torus
$T^4$ that it is thought as a trivial elliptic fibration of a 
two-dimensional fiber torus $T^2_f$ over another two-dimensional base torus
$T^2_b$. The fiber torus has real coordinates $(x_0, x_1)$ while the base 
torus has coordinates $(x_2, x_3)$. We choose a complex structure
on the fiber torus with complex coordinates $(z, \bar z)$ and a complex 
structure
on the base with coordinates $(u, \bar u)$. Tangent directions to the fiber are 
indicated by $\parallel$, while directions tangent to the base with $\perp$
. With this notation, introducing the auxiliary variable $E_{\perp}$, 
the partition function can be written as a Gaussian integral over 
$E_{\perp}$:
\bea
Z = \int \exp\left[- \frac{N}{2\lambda}\sum_{\parallel \perp}\int 
Tr \left(F_{\parallel}^2+ E_{\perp}^2-2i E_{\perp} F_{\perp}
+F_{\parallel \perp}^2\right) d^4x\right] DA_{\parallel} DA_{\perp} DE_{\perp}.~~~~
\eea
The integration over the components of the connection transverse
to the fiber, $A_{\perp}$, is Gaussian and can be performed explicitly,
giving as a result the effective action for $E_{\perp}$ and 
$A_{\parallel}$~\cite{B}.
From now on, $A_{\parallel}$ will be indicated simply as $A$ and $E_{\perp}$
as $E$ unless stated otherwise. This gives for the partition function:
\bea
Z&=&\int \exp\left[- \frac{N}{2\lambda}\int Tr \left(F_{A}^2+ 
(\partial_{\perp}A)^2+ E^2+\cdots\right)
 d^4x\right] \times
\nonumber \\
&&\hphantom{\int}
\times Det\left[-\Delta_A-i \, ad_E\right]^{-1} DA DE \, ,
\eea
where the dots indicate non-local terms, some of which will be made to
vanish later on, by an appropriate choice of the gauge-fixing
condition, and the sum over the appropriate space-time indices is
understood.  $\Delta_A$ is the two-dimensional scalar Laplacian in the
background of the connection $A$ and $ad_E$ the adjoint action of the
Lie algebra valued field $E$.  At this point, we make a change of
variables, that allows us to embed $T^* \cal {A}$ into the QCD
functional integral. We simply set
\bea
E_{\perp}=\epsilon_{\perp \parallel} F_{\parallel}(A_D)\,,
\eea
where $\epsilon_{\perp \parallel}$ is the rank four normalized antisymmetric
tensor.
We call this change of variables a fiberwise duality 
transformation, because it has the structure of a duality transformation 
restricted to the fiber~\cite{B}. Indeed the complete duality transformation in
four dimensions~\cite{Ha,MM} would involve solving the Bianchi identities 
for the dual field strength:
\bea
\left[\partial_{\mu}+i A^{D}_{\mu}, \tilde{K}_{\mu \nu}\right]=0\,,
\eea
that imply
\bea
K_{\mu \nu}=F_{\mu \nu}(A_D)\,,
\eea
up to some field-strength copy problem~\cite{Halpern,SS}.
To implement this change of variables in the functional integral
we employ the resolution of the identity, by means of the  Faddeev-Popov
trick:
\bea
1=\Delta_D(E)  \int \delta[E-F(A_D)] DA_D \, ,
\eea
where $F(A_D)$ is a dual curvature two-form associated to the dual connection
one-form $A_D$.
It is convenient for our purposes to decompose $A_D$ into $A$
and an arbitrary one-form $\Psi$:
\bea
A_D=A+\Psi \, .
\eea
Correspondingly, the Faddeev-Popov trick becomes, after shifting the $A_D$
integration~by~$A$:
\bea
1=\Delta_D(E,A)  \int \delta[E-F(A+\Psi)] D\Psi \, .
\eea
We may consider the fields $(A, \Psi)$ as the coordinates of $T^* \cal 
{A}$,
the cotangent space of unitary connections, $A$, on the fiber torus. $(A, \Psi)
$ are four-dimensional fields as functions on space-time,
but they belong to $T^* \cal {A}$ fiberwise.
The QCD functional integral then becomes:
\bea
Z&=&\int \exp\left[- \frac{N}{2\lambda}\int Tr \left(F_{A}^2+ 
(\partial_{\perp}A)^2+ E^2+\cdots\right)
 d^4x\right] \times 
\nonumber \\
&& \hphantom{\int}
\times Det\left[-\Delta_A- i \, ad_E\right]^{-1} \Delta_D(A,\Psi) 
 \delta[E-F_{A+\Psi}] DA D\Psi DE \,.
\eea
The $DE$ integration can now be performed by a convenient choice of
the gauge, that has also the advantage of eliminating some of the
non-local terms in the functional integral that were indicated by the
dots.  The gauge choice is~\cite{B}:
\bea
d^*A+E=0 \,,
\eea
where $d^*$ indicates the two-dimensional divergence of $A$.
Inserting this gauge condition and the corresponding Faddeev-Popov 
determinant $\Delta_{FP}$ and performing the $DE$ integration we get:
\bea
Z&=&\int \exp\left[- \frac{N}{2\lambda} \int Tr\left( F_{A}^2+(d^* A)^2+
 (\partial_{\perp}A)
^2+\cdots\right) d^4x\right] \times 
\nonumber \\
&& \hphantom{\int}
\times Det[-\Delta_A- i \, ad_{F_{A+\Psi}}]^{-1} \Delta_D(A,\Psi) 
\Delta_{FP} 
\times 
\nonumber \\
&&  \hphantom{\int}
\times \delta[d^*A+F_{A+\Psi}] DA D\Psi \, .
\eea
This gives the desired embedding of  
$T^* {\cal {A}}/G$ fiberwise in the 
functional integral.

Analytically our problem is, apparently, as difficult as the original one.
Instead of integrating over the connection one-form in four dimensions
(minus a gauge-fixing condition) we integrate over $T^* \cal {A}$
fiberwise (minus a gauge-fixing condition). After gauge-fixing, in both
cases, the effective action is, in general, non-local.

Our main step will be, nevertheless, to use algebraic geometry to give
the quotient, 
$T^* {\cal {A}}/G$, of $T^*\cal{A}$ by the gauge group $G=SU(N)
$, an explicit meaning as a moduli space.
The fact that this moduli space will turn out to be an integrable system
will be essential for our approach to the large $N$ limit.
Had we tried to construct directly, in some sense, the moduli space of four-
dimensional bundles, without integrating out the transverse polarizations of
the connection, we would not have obtained such integrability properties.

To understand the moduli space, we must understand the structure of the 
gauge orbits in $T^* \cal {A}$.
The gauge group acts on the fields $A$, $\Psi$ by hamiltonian vector fields
with respect to the three symplectic forms~\cite{H1}.
The corresponding moment maps are:
\bea
\mu^0&=&F_A-i[\Psi,\Psi]\,,
 \nonumber \\
\mu&=&\bar{\partial}_A \psi \,,
\nonumber \\
\bar{\mu}&=&\partial_A \bar{\psi} \, ,
\eea
where in this formula $\psi$ and $\bar{\psi}$ refer to the $z$ and $\bar z$
components of the one-form  $\Psi=\psi+\bar{\psi}$.
General principles of hyper-Kahler geometry require, for the quotient
under the action of the gauge group to be separable and hyper-Kahler,
the moment map to be central. For $SU(N)$ on a compact Riemann surface
this forces the moment map to vanish, unless the structure group of the
cotangent bundle is reducible. This is a too restrictive constraint in the
functional integral. However, we can
consider bundles with a parabolic flag structure and weights on an arbitrary 
divisor 
$D$~\cite{MS,M1}.
Set in another way, we put coadjoint orbits on the given divisor
associated to the parabolic structure~\cite{Ale}.
This amounts to add, to the three moment
maps for the gauge connection and the Higgs current, delta-like
singularities on the divisor.
A collective field for the parabolic Higgs bundles can then be introduced in 
the functional integral by means of the resolution of the identity~\cite{B1}:
\bea
1&=&\lim_{|D| \rightarrow \infty}
 \int \delta\left(\mu^0-\frac{1}{|D|}\sum_p \mu^{0}_{p} \delta_p\right) \times 
\nonumber \\
&& \hphantom{\lim_{|D| \rightarrow \infty}\int}
\times \delta\left(\mu-\frac{1}{|D|}\sum_p \mu_{p} \delta_p\right) 
 \delta\left(\bar{\mu}-\frac{1}{|D|}\sum_p \bar{\mu}_{p} \delta_p\right) \times  
\nonumber \\[5pt]
&&\hphantom{\lim_{|D| \rightarrow \infty}\int}
  \times \prod_p d\mu_p^0 d\mu_p d\bar{\mu}_p d^2z_p \,
J[\mu_0,\mu,\bar{\mu}] \,,
\label{14}
\eea
where $\delta_p$ is the two-dimensional delta-function localized at $z_p$ 
and $J$ a compensating Jacobian determinant.
This collective field for the moment maps is the analogue, in our 
gauge-theoretic setting, of the collective field for the density of the 
eigenvalues, introduced long ago for solving the large $N$ limit of matrix 
models~\cite{J}. All the variables in Eq.~(\ref{14}) are functions of the 
coordinates $(u, \bar{u})$ on the base torus. We will leave implicit
this functional dependence in all the following formulae.
The levels of the moment maps are dense in the sense of the
distributions, as the divisor
gets larger and larger.
$\mu_p^0$ is a traceless hermitian matrix while $\mu_p$ and $\bar{\mu}_p$ live
in the Lie algebra $sl(N)$. There is still the freedom of making $SU(N)$
gauge transformations. This was implemented, in the previous discussion,
by the choice of the gauge-fixing.

A deeper justification for considering moment maps valued in the
distributions relies on the following facts.  In two dimensions (or
fiberwise, as in our case) there is a correspondence between unitary
bundles and holomorphic ones~\cite{A}. In general, to study the moduli
problem, it is more convenient the holomorphic language, since
geometric invariant theory tells us when a moduli space exists as a
separable manifold.  A necessary and often sufficient condition is
stability in the sense of Mumford~\cite{NS}.  Now, in a remarkable
theorem, Narasimhan and Seshadri~\cite{NS} proved that semi-stable
holomorphic bundles on a compact Riemann surface arise as unitary
projective representation of the fundamental group of the surface (or
unitary representations of the once-punctured surface). Stable bundles
correspond to irreducible representations. In addition, the moduli of
non-stable bundles cannot be separable in general~\cite{NS}.  The
Narasimhan-Seshadri result has been extended to the unitary
non-compact case by Metha and Seshadri~\cite{MS}, who proved that
stable holomorphic bundles with a parabolic flag structure (of
parabolic degree zero) arise as unitary representation of the
fundamental group of a compact Riemann surface with marked points and
fixed monodromy conjugacy classes around the marked points. There is,
moreover, a correspondence between the weights of the parabolic
structure and the unitary eigenvalues of the monodromy, and between
the flag structure and the eigenspaces of the eigenvalues of the
monodromy as well. More simply, the moduli of these bundles can be
described, in the language of moment maps and symplectic reduction, as
a quotient of the space of unitary connections with coadjoint orbits
under the action of the gauge group~\cite{Ale}. The computation of the
analytic torsion in $YM_2$ fits into this symplectic
framework~\cite{W}.  Later on, an analogous computation, in the
hyper-Kahler case of Hitchin bundles, will play some role in this
paper.  Hitchin extended the Narasimhan-Seshadri result in another
direction. He answered the question to what correspond non-unitary
representations of the fundamental group of a compact
surface~\cite{H1}. The answer is stable (semi-stable) Higgs bundles.
Given a decomposition of a non-unitary flat connection into a unitary one plus a one-form valued in the complement of the compact
generators, it is possible to choose the one-form in such a way that it is 
harmonic with respect to the unitary connection~\cite{S}. The
corresponding decomposition of the non-hermitian curvature furnishes
the three moment maps for the action of the compact subgroup.
Finally, Hitchin result has been extended to the non-compact
non-unitary case by several other authors~\cite{Y,K,M1,S}. The moduli
space of parabolic $K(D)$ pairs is the moduli of holomorphic bundles
with a parabolic structure and a holomorphic endomorphism with
parabolic residue. It is a Kahler, but not hyper-Kahler manifold, of
dimension (in the traceless case)~\cite{Y}:
\bea
dim {\cal{P}}^0 _{\alpha}=\left(2g-2+\mid D \mid\right)\left(N^2-1\right) \, .
\label{15}\eea
It contains as a closed subspace the moduli of Higgs bundles, for which the 
endomorphism has nilpotent residue. 
\pagebreak[3]
The last one is a hyper-Kahler manifold,
of dimension (in the traceless case)~\cite{Y,K}:
\bea
dim {\cal{N_{\alpha}}}^0=(2g-2)(N^2-1)+2 \sum_{p \in D} f_p \, ,
\eea
where $f_p=\frac{1}{2}(N^2-\sum_{i} m_i(p)^2)$ and $m_i$ is the multiplicity of
the weight $\alpha_i$ associated to the flag structure.

A consequence of this discussion is the following philosophy for
approaching the large $N$ limit.  Sometimes to solve a problem it may
be convenient to embed it into a family.  In this case, we want to
find the master field by embedding the fluctuations of the theory into
a dense family. In fact we consider fluctuations modulo gauge
equivalence. It is precisely at this point, after gauging away the
gauge group, that the separability of the moduli space requires
introducing representations of the fundamental group fiberwise,
something very natural in a gauge theory.  Since the theory is local,
the natural objects are the representations of the fundamental group
associated to an arbitrary divisor. This forces the moment maps to be
distributions concentrated on the divisor. The requirement of having a
dense separable family of gauge-invariant fluctuations has lead us to
a very weak topology. This is unavoidable, but the advantage is the
integrability of the moduli space.

As a check, let us come back to the functional integral and count,
somehow naively, the dimension of the local moduli space. For each
point of the divisor we have a traceless hermitian plus a 
traceless matrix and its complex conjugate, the moment maps, minus a
traceless hermitian gauge condition. Following this counting the
number of local moduli is $\mid D \mid (N^2-1)$ complex parameters. We
should then take into account also the global properties of the torus,
but it turns out that on a torus there is no extra contribution,
according to Eq.(\ref{15}).  The hermitian moment map $\mu^0$ determines a
collection of weights via its eigenvalues modulo $2 \pi$ and a flag
structure via its eigenspaces.  A holomorphic sheaf is obtained
gauging away the field $A$ by means of a complexified gauge
transformation (we should stress that complexified gauge
transformations are simply a convenient change of variables to
describe the moduli in the holomorphic language, not gauge
transformations in the functional integral).  In this gauge $\Psi_z$ is
holomorphic with a certain residue. Generically the residue is
conjugated, in the complexified gauge group, to a parabolic element.
We have thus a parabolic sheaf together with a parabolic endomorphism
$\Psi_z$ .The corresponding moduli space is the one of parabolic $K(D)$
pairs.  If the residue of $\Psi_z$ is chosen to be nilpotent the
corresponding moduli space is the one of parabolic Higgs
bundles~\cite{Y,K}, that is hyper-Kahler.  The inconvenient for
controlling the large $N$ limit is, however, that the collective field
of $K(D)$ pairs or of Higgs bundles has still a number of components
of order $N^2$.  As mentioned in the introduction, a crucial reduction
in the entropy is obtained passing to a collective field associated to
the Hitchin fibration.
\pagebreak[3]

According to Hitchin, by a complexified gauge transformation, $\Psi_z$
can be made holomorphic (with possibly a parabolic or nilpotent
residue on the parabolic divisor).  The equation
\bea
Det( \lambda1 - \Psi_z)=0
\eea
determines a spectral cover as a polynomial equation in $\lambda$
with, as coefficients, $N$ symmetric invariant polynomials in $\Psi_z$,
that are $N$ (meromorphic)~\cite{H3,N} differentials. The spectral
cover is branched at points where the characteristic polynomial has
multiple roots.  Since $\Psi_z$ can always be put in triangular form by
an extra holomorphic gauge transformation, the eigenspace
associated to the first eigenvalue $\lambda^1_z$ is one dimensional
and depends on $z$. This defines a holomorphic line bundle.  We are
thus given a holomorphic line bundle $L$ and a spectral covering
$\pi^{-1}$.  Hitchin shows that $\Psi_z$ can be recovered by $L$ and
$\pi^{-1}$. Given a section $\sigma$ of $L$, the sections of the Higgs
bundle in the holomorphic gauge can be recovered as the direct image
bundle $\sigma(\pi^{-1})$~\cite{H3,N,M1,M2}.  Then the multiplicative
action of $\lambda^1_z$, thought as a differential on the covering, on
the section $\sigma$ of the line bundle $L$, determines $\Psi_z$ as an
endomorphism of the direct image bundle~\cite{H3}.  This gives $\Psi$
in the holomorphic gauge.  To go to a unitary gauge, by making a
complexified gauge transformation, we determine $A$ as a solution of
the hermitian moment map equation. This leaves still the freedom of
making unitary gauge transformations and leaves the choice of an
arbitrary hermitian gauge fixing.  We have thus recovered the $K(D)$
or Higgs bundle from $L$ and the spectral cover.  Let us observe that
the sheaf of sections of a holomorphic line bundle, that is equivalent
to the line bundle itself, is determined as the kernel of the (0,1)
part of an abelian $U(1)$ connection, $a=a_z dz+a_{\bar z}
d\bar{z}$,~\cite{A}.  The spectral covers instead are determined by
$N$ meromorphic $q$-differentials, $a_q=a_{z^q} dz^q$.  The Hitchin
data are therefore those of a vector-like model of complex dimension
$N+1$ ($N$ in the traceless case). Yet, the dimension of the moduli
space of the Higgs bundles and the one of the line bundles on the
spectral coverings, including the moduli of the coverings, are exactly
the same.  We are now ready to implement this change of variables in
the functional integral. We first describe the change of variables at
formal level and afterward we exploit the detailed structure of the
Jacobian determinant.  The QCD partition function can be rewritten as:
\bea
&&Z = \int \exp\left[- \Gamma(A,\Psi)+ 
 \log \frac{\partial\left(A,\bar{A},\Psi,\bar{\Psi}\right)}
{\partial\left(a,\bar{a},a_{q},\bar{a}_{q}\right)}\right] 
 Da D\bar{a} Da_{q} D\bar{a}_{q}\,,
\eea
that is, formally, the result looked forward.
We want now to give the Jacobian of the change of variables a precise 
meaning. We describe the computations in a rather sketchy way. We think the
functional integral to be stratified by the levels of the 
\pagebreak[3]
moment maps. This is done inserting the obvious resolution of the identity
into the functional integral, 
corresponding to the levels, to the 
over-complete resolution of the levels in terms of coadjoint 
orbits and to 
the gauge-fixing:
\bea
Z &=& \lim_{|D| \rightarrow \infty }
\int \exp[- \Gamma(A,\Psi)] \times 
\nonumber\\
&& \hphantom{\lim_{|D| \rightarrow \infty }\int}
\times \delta\left(\mu^0-F_A+i[\Psi,\Psi]\right) 
\delta\left(\mu-\bar{\partial}_A \psi\right) 
\delta\left(\bar{\mu}-\partial_A \bar{\psi}\right) \times 
\nonumber\\
&&\hphantom{\lim_{|D| \rightarrow \infty }\int}
\times  \delta\left(\mu^0-\frac{1}{|D|}\sum_p \mu^{0}_{p} \delta_p\right) 
\delta\left(\mu-\frac{1}{|D|}\sum_p \mu_{p} \delta_p\right)
\delta\left(\bar{\mu}-\frac{1}{|D|}\sum_p \bar{\mu}_{p} \delta_p\right)  
\times 
\nonumber \\
&&\hphantom{\lim_{|D| \rightarrow \infty }\int}
 \times J\left[\mu_0,\mu,\bar{\mu}\right] \,
\Delta_{FP} \,  \delta\left[d^*A+F_{A+\Psi}\right]
D\mu^0 D\mu D\bar{\mu} \times 
\nonumber\\[.5ex]
&&\hphantom{\lim_{|D| \rightarrow \infty }\int}
 \times \prod_p d\mu_p^0 d\mu_p d\bar{\mu}_p d^2z_p
DA_z D\Psi_z DA_{\bar z} D\Psi_{\bar z} \, .
\eea
The integration over $ DA_z D\Psi_z DA_{\bar z} D\Psi_{\bar z} $ gives a
combination of functional 
determinants, $T$,(the analogue of the analytic torsion~\cite{W}) times 
the Kahler (hyper-Kahler) volume, 
$\wedge^{max} \omega_{\vec{\mu}_{p}}$, associated to the orbits with given 
level of the moment map, $\vec{\mu}_{p}$ (we have grouped the three moment 
maps into a vector).
We now insert the collective field, $\rho$, associated to the divisor $D$ and 
the corresponding Jacobian: 
\bea
Z &=& \int \exp[- \Gamma(A,\Psi)]\,
\frac{J[\vec{\mu}(\rho)]}{J(\rho)} \, T \wedge^{max} 
\omega_{\vec{\mu}(\rho)} 
\, D\rho \, ,
\eea
where $ \lim_{|D| \rightarrow \infty } \int
\delta\left(\rho-\frac{1}{|D|}\sum_p \delta_p\right) \prod_p d^2z_p
\,J(\rho)=1$ and the dependence from the divisor is now expressed
through the collective field $\rho$ (this is certainly possible if the
moment maps are a symmetric function of the points of the divisor; in
particular if $\vec{\mu}_{p}=\vec{\mu}$ for all $p$, as it is expected
for a translationally invariant ground state).  Then, we change
variables to the Hitchin fibration in terms of the the symplectic
volume form $\wedge^{max} \omega_{\rho}$ associated to the symplectic
structure on the cotangent space of rank $1$ bundles on the spectral
covering:
\bea
Z &=& \int \exp[- \Gamma(A,\Psi)] \times 
\nonumber\\
&&\hphantom{\int} 
 \times
\frac{J[\vec{\mu}(\rho)]}{J(\rho)}  \,T \, \frac{\wedge^{max} 
\omega_{\vec{\mu}(\rho)}}{\wedge^{max} \omega_{\rho}} \times{\wedge^{max} 
\omega_{\rho}}
 \, D\rho \, ,
\label{21}
\eea
where $\omega_{\rho}$ can be chosen to be a symplectic form on rank $1$
Higgs bundles on the spectral covers
$\pi^{-1}(T^2_f)$, with coordinates $(a(\rho), \lambda^1(\rho))$ :
\bea
\omega_{\rho} &=& \int_{\pi^{-1}(T^2_f)} \delta a(\rho) \wedge  \delta \bar{a}
(\rho) + 
\delta \lambda^{1}(\rho) \wedge
\delta \bar{\lambda}^{1}(\rho) \,.
\eea
Finally, we represent $\wedge^{max} \omega_{\rho}$ as the $U(1)$ analytic 
torsion \cite{W} on the coverings times the volume of the coverings:
\bea
\wedge^{max} \omega _{\rho} &=& \delta\left.\left(F_{a(\rho)}\right) 
\right|_{\pi^{-1}(T^2_f)} \delta[d^*a(\rho)]
\Delta_{FP}(a(\rho)) Da(\rho) 
D\bar{a}(\rho) \times 
\nonumber \\
&&\times
\frac{\partial(\lambda(\rho), \bar{\lambda}(\rho))}{\partial\left(a_{q}(\rho),
\bar{a}_{q}(\rho)\right)}
 Da_{q}(\rho) D\bar{a}_{q}(\rho) \, ,
\label{23}
\eea
where, in the volume form for the coverings, we have changed
variables, from the eigenvalues $\lambda$ of $\Psi_z$, to the
meromorphic differentials $a_{z^q}$ ($=Tr[{\Psi_z}^q]$), holomorphic
away from the divisor $D$. Notice that the entropy, associated to the
measure in Eq.~(\ref{23}), is of order $N$, since only $N$ fields,
$a_{z^q}$, living on $T^2_f$ and one, $a$, living on $\pi^{-1}(T^2_f)$
(that is a $N$-sheeted covering) are integrated. Actually we should
take into account also the integration over the collective field of
the divisor $\rho$ in Eq.~(\ref{21}), but this is still a contribution of
order one.  This completes our sketchy evaluation of the Jacobian
determinant to the collective field of the Hitchin fibration.

\section{Conclusions}

To summarize, we represent the QCD partition function, by means of a
fiberwise duality transformation on the base torus, as a 
functional integral on $T^*\cal{A}$ fiberwise:
\bea
Z&=& \int \exp\left[- \frac{N}{2\lambda} \int Tr\left( F_{A}^2+(d^* A)^2+
 (\partial_{\perp}A)
^2\right)+ (\hbox{non-local terms})\, d^4x\right] \times
\nonumber \\
&&\hphantom{\int}
 \times Det\left[-\Delta_A-i \, ad_{F_{A+\Psi}}\right]^{-1}
\Delta_D(A,\Psi)\times\nonumber\\[3pt]
&&\hphantom{\int}\times\Delta_{FP} \delta[d^*A+F_{A+\Psi}]
DA_z D\Psi_z DA_{\bar z} D\Psi_{\bar z}= 
\nonumber \\[3pt]
&=&\int \exp[- \Gamma(A,\Psi)]
DA_z D\Psi_z DA_{\bar z} D\Psi_{\bar z} \, .
\eea
Then, we choose a dense subset of $T^*\cal{A}$ (dense in the large $N$
limit), 
that admits a separable Kahler (hyper-Kahler) quotient under the action 
of the
gauge group. This dense subset turns out to be the inductive limit of the 
sheaves of $ K(D) $ pairs (Higgs bundles) with arbitrary parabolic divisor 
$D$.
Thereafter, on this dense subset, we change variables from
$(A_z, \Psi_z, A_{\bar z}, \Psi_{\bar z})$ to $(a_z,a_{z^q},a_{\bar z},
a_{\bar{z}^q})$, where $(a_z, a_{\bar{z}})$ is a $U(1)$ connection on a $N$-
sheeted covering and $(a_{z^q}; 
q=1,...,N)$
are $N$ gauge-invariant meromorphic $q$-differentials, holomorphic away
from $D$, fiberwise over the base torus. 

Hence, since our theory is now labelled by a $U(1)$ connection plus a set of
$N$-chiral and anti-chiral fields of increasing spin, up to $N$, it turns out
that the effective action of QCD, in the large $N$ limit, reads:
\bea
 \Gamma_{eff}(A,\Psi)= \Gamma(A,\Psi)+ \log \frac
{\partial(a,\bar{a},a_q,\bar{a}_q)}
{\partial(A,\bar{A},\Psi,\bar{\Psi})}    \, .
\eea
The effective action $\Gamma_{eff}$, as a functional 
of a parabolic Higgs bundle, need not to be finite, because of the 
singularities of the connection and Higgs field on the parabolic divisor. 
However, this is not unexpected. The changes of variable, that we 
performed in the functional integral to get the effective action, are by no
means a substitute of regularization and renormalization. In this respect we 
are in the same situation as for the Migdal-Makeenko equation~\cite{MiMa}, 
that needs to be regularized and renormalized.
What is needed here is a regularization of the theory that respects as 
most as possible the moduli structure of the parabolic Higgs bundles
and of the Hitchin fibration, that is, the gauge structure of the theory.
Surprisingly, such regularization exists and it consists in analytically
continuing the fields $(A,\Psi)$ from Euclidean to Minkowskian
space (sic!)~\cite{P} (p.186). This gives every pole, in the $(x_+ , x_- )$ 
Minkowskian coordinates, a $i \epsilon$ part, moving away the singularity from
the real integration path. The divergences are then obtained in the $ \epsilon
 \rightarrow 0 $ limit and $ \epsilon $ is our regularization
parameter.

An important point is that the master field should be gauge equivalent to a 
constant~\cite{W2}.
In the present approach, instead, it is described as a collective field of 
distributions. Yet, it is conceivable that this collective field
converges to a constant, in the large $N$ limit, in the sense of the
distributions. In fact, the weak topology is often the strongest 
topology that may be introduced in field theory, as
the asymptotic weak convergence of interacting fields to free fields,
in scattering theory, shows.

\acknowledgments

We would like to thank Corrado De Concini for several enlightening
discussions on the Hitchin fibration.

\end{document}